\documentclass[aps,prl,reprint, floatfix, 10pt, showpacs,groupaddress]{revtex4-1}  

\newcommand{\cm}{\mbox{ cm}}

\newcommand{\Rb}{\ensuremath{^{87}}Rb }
\newcommand{\ket}[1]{\ensuremath{\left| #1 \right>}}

\usepackage{graphicx}
\usepackage{epstopdf}
\usepackage{bbold}
\usepackage{comment}
\usepackage{color}
\usepackage{soul}
\usepackage[abs]{overpic}
\usepackage{amsmath,amsfonts,amssymb}
\usepackage{lipsum}
\usepackage{hyperref}
\usepackage{setspace}
\usepackage{siunitx}
\usepackage{afterpage}
\usepackage{braket}
\usepackage{bm}
\usepackage[bottom]{footmisc}

\DeclareSIUnit\torr{Torr}

\makeatletter
\AtBeginDocument{\let\LS@rot\@undefined}
\makeatother

\graphicspath{{Figures/}}
\begin{document}


%
\author{Ronen Chriki}
\thanks{These authors contributed equally}
\author{Slava Smartsev}
\thanks{These authors contributed equally}
\author{David Eger}
\author{Ofer Firstenberg}
\author{Nir Davidson}
\email{nir.davidson@weizmann.ac.il}
\affiliation{Department of Physics of Complex Systems, Weizmann Institute of Science, Rehovot 7610001, Israel }

\begin{abstract}
Partially coherent light is abundant in many physical systems, and its propagation properties are well understood. Here we extend current theory of propagation of partially coherent light beams to the field of coherent diffusion. Based on a unique four-wave mixing scheme of electro-magnetically induced transparency, an optical speckle pattern is coupled to diffusing atoms in a warm vapor. The spatial coherence propagation properties of light speckles is studied experimentally under diffusion, and is compared to the familiar spatial coherence of speckles under diffraction. An analytic model explaining the results is presented, based on a diffusion analogue of the famous Van Cittert-Zernike theorem. 
\end{abstract}

\title[]{Coherent diffusion of partial spatial coherence}
\maketitle


Spatial correlations exist in many different physical systems, and the study of their origin and evolution is one of the primary roles of statistical physics. In optics, the propagation of spatial coherence of partially coherent light sources has attracted much attention ever since the early days of modern optics~\cite{goodman2015statistical, mandel1995optical}. One of the most prominent theorems in optics is the Van Cittert-Zernike (VCZ) theorem~\cite{van1934wahrscheinliche, zernike1938concept}, which describes the spatial coherence far from a spatially incoherent source. This theorem shows a Fourier relation between the intensity distribution on the surface of a spatially incoherent source and the spatial coherence far from it. Physically, this implies that the boundaries of the source dictate the coherence properties of the illuminated light far from the source. This property of the spatial coherence has been famously exploited in Michelson's stellar interferometer to measure the size of distant radiation sources~\cite{michelson1921measurement, hariharan2003optical}. Hanbury Brown and Twiss later showed that similar stellar information can be extracted by measuring intensity correlations~\cite{brown1956test, brown1956correlation, brown1974intensity}. While the classical theorems describe the spatial coherence far from the source (VCZ region), more recent studies consider short propagation distances, in the region near the source (deep Fresnel region)~\cite{cerbino2007correlations, gatti2008three}, and show that therein the spatial coherence is propagation invariant~\cite{ohtsuka1981non, turunen1991propagation, giglio2000space}. 

The concepts and theorems derived in linear optics were later extended to interacting photons~\cite{HBT_interactingPhotons}, as well as to atomic and condensed matter systems, demonstrating partial spatial coherence of electrons~\cite{HBTelectrons1, HBTelectrons2, HBTelectrons3} and cold atoms~\cite{HBTColdAtoms, HBT_acrossBEC, HBTFermionsAndBosons, HBTMatterWaves}. The underlying assumption in all these systems is that they only exhibit ballistic transport, while any diffusive transport is negligible. Although this assumption is well justified in many cases, it does not always hold, and often diffusive transport must be taken into consideration~\cite{HBTcoldAtomsDiffusion}. Here we thoroughly investigate coherent diffusion of spatial correlations (spatial coherence) of partially coherent fields, theoretically and experimentally, and compare between diffraction and coherent diffusion of partial coherence. 

The comparison between these two distinct physical mechanisms is based on the mathematical similarity between their governing equations,
\begin{equation}
\begin{aligned}
    &\text{coherent diffusion: }
    \frac{\partial \phi\left(\mathbf{r},t\right)}{\partial t} = D\triangledown^2_\mathbf{r}\phi\left(\mathbf{r}, t\right)\\
    &\text{paraxial diffraction: }
    \frac{\partial \phi\left(\mathbf{r},t\right)}{\partial z} = \frac{i\lambda}{4\pi}\triangledown^2_\mathbf{r}\phi\left(\mathbf{r}, t\right),
    \label{eq:DiffusionVsDiffraction}
\end{aligned}
\end{equation}
where $\phi$ is a complex field, $\mathbf{r}$ is the transverse coordinate, $t$ is time, $z$ propagation distance, $D$ is the diffusion coefficient and $\lambda$ the wavelength. Equation~(\ref{eq:DiffusionVsDiffraction}) presents the familiar diffusion equation (Fick\textsc{\char13}s second law of diffusion); but as opposed to the traditional text-book examples of diffusion, such as diffusion of heat or concentration, $\phi$ here is complex valued, rather than real~\cite{xiao2008slow, firstenberg2008theory,  grebenkov2007nmr}. 

Clearly the two equations above are identical under the transformation $D\to i\lambda /4\pi$, and accordingly diffraction can be considered as diffusion in imaginary time~\cite{firstenberg2013colloquium, firstenberg2009elimination}. This mathematical similarity implies exciting physical analogies, where various well-known optical phenomena find their natural analogues in diffusion of complex vector fields~\cite{pugatch2007topological, shuker2008storing, firstenberg2009elimination, firstenberg2010self}. For example, optical vortices are topologically protected in both diffusion and diffraction~\cite{pugatch2007topological}. Needless to say, although a mathematical similarity exists between diffraction and diffusion, there are many differences between the two. Probably the most prominent difference is related to dissipation: Diffraction is an energy conserving phenomena, and can therefore be reversed, whereas diffusion is a dissipative process.

\begin{figure}
    \centering
    \includegraphics{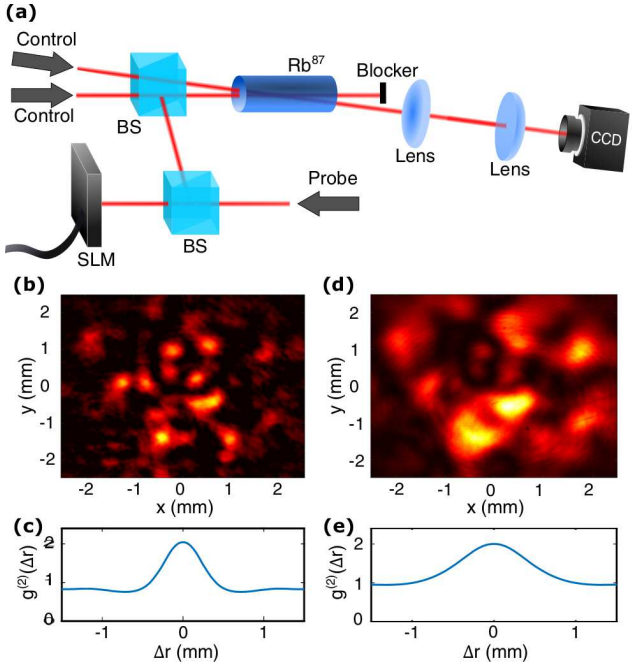}
	\caption{Experiment and representative results. (a) Experimental arrangement for diffusion of speckle fields. BS - beam splitter; SLM - spatial light modulator. (b, d) Detected intensity distribution of the vapor cell, at large and small detuning (short and long diffusion time). (c, e) Corresponding autocorrelation of the detected speckle pattern.}
	\label{fig:Experimental_arrangement}
\end{figure}

{\it Experimental results.---} The experimental arrangement used to characterize the diffusion of partially coherent fields is illustrated in Fig.~\ref{fig:Experimental_arrangement}(a). We use \Rb which diffuses in \SI{10}{\torr} of $\text{N}_2$ buffer gas, rendering a diffusion coefficient of $D=\SI[separate-uncertainty = true, multi-part-units=single]{9.7(5)}{\cm\squared\per\second}$. The vapor cell is illuminated by two spatially overlapping 'control' beams, which are separated by a slight angle; and by a third weak 'probe' beam, which is oriented along one of the control beams. consequently, a fourth beam, denoted as 'signal', is generated in a four-wave mixing process along the orientation of the second control beam. We set the optical frequencies of the probe and control beams such that they couple, respectively, the lower and upper hyperfine states $\ket{1}=\ket{5S_{1/2}; F=1,2; m=0}$ to the excited states $\ket{2}=\ket{5P_{1/2}; F'=1,2; m=1}$ and $\ket{3}=\ket{5P_{1/2}, F'=1,2; m=-1}$ of the $D_1$ transition. The incoming probe beam $E_\text{in}(\mathbf{r})$ is shaped using a spatial light modulator. The outgoing signal $E_\text{out}(\mathbf{r})$ is imaged onto a CCD camera, and we use digital Fourier filtering to improve the signal-to-noise ratio. Further details regarding the experimental arrangement are given in the SI~\cite{Supplementary}, whereas full characterization and analysis of the generation process are described in  Ref.~\cite{smartsev2017continuous}.

Using the Fourier transformation for the transverse coordinates $\mathbf{r}=(x,y)$, $\Tilde{E}(\mathbf{q}) = \int \frac{d^2\mathbf{r}}{2\pi}  E(\mathbf{r})e^{-i\mathbf{q}\cdot\mathbf{r}}$ and under the assumptions of weak EIT and confined spatial frequencies $q^2=|\mathbf{q}|^2\ll|\gamma_{2p} + \Gamma|/D$, it can be shown that \cite{smartsev2017continuous, Supplementary}
\begin{equation}
\label{eq:DiffusionInFourier}
    \Tilde{E}_S(\mathbf{q}) = \frac{\Tilde{E}_S(\mathbf{q}=0)}{\Tilde{E}_\text{in}(\mathbf{q}=0)}\cdot\Tilde{E}_\text{in}(\mathbf{q})e^{-D\tau q^2},
\end{equation}
where $\tau$ is the group delay of the signal~\cite{Supplementary}
\begin{equation}
\label{eq:DiffusionTime2}
    \tau = \frac{\tau_\infty}{1+\left(\Delta_{\text{2p}}\tau_\infty\right)^2},
\end{equation}
$\Delta_{\text{2p}}$ being the two-photon frequency detuning and $\tau_\infty$ the maximal diffusion time that can be achieved, $\tau_\infty\equiv1/(\gamma_{\text{2p}}+\Gamma)$. Here $\gamma_{\text{2p}}$ is the decoherence rate of the two-photon transition and $\Gamma$ denotes the power broadening, $\Gamma=\Omega^2/(\gamma_\text{1p}-i\Delta_\text{1p})$, $\Omega$ being the Rabi frequency of the control beams, $\gamma_\text{1p}$ the one-photon linewidth  and $\Delta_\text{1p}$ the one-photon frequency detuning. In our experiments, $\tau_\infty\approx\SI{81}{\micro\second}$.

The propagator $e^{-D\tau q^2}$ in Fourier space implies diffusion in real space. It follows from Eq.~(\ref{eq:DiffusionInFourier}) that  a structured probe beam in our system can continuously generate a signal which underwent diffusion for an effective temporal duration $\tau$. Figure~\ref{fig:Experimental_arrangement}(b) presents a representative retrieved signal for an input Gaussian speckle field, under large detuning $\Delta_{\text{2p}}$ (i.e. short diffusion time $\tau=\SI{4}{\micro\second}$), and Fig.~\ref{fig:Experimental_arrangement}(d) shows the retrieved signal for the same speckle pattern, under small detuning (i.e. long diffusion time $\tau=\SI{65}{\micro\second}$). Figures \ref{fig:Experimental_arrangement}(c,e) show the autocorrelation of the retrieved intensity patterns. Based on such measurements and for various values of $\Delta_{\text{2p}}$, we can study the effect of diffusion on the coherence of speckle fields. 

\begin{figure}
    \centering
    \includegraphics{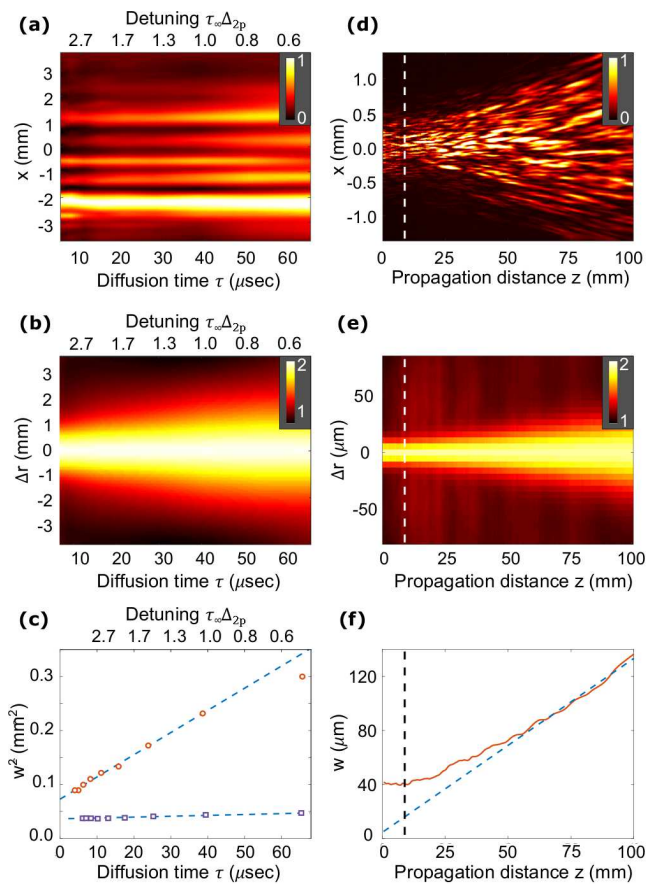}
	\caption{Experimental comparison between diffusion (a-c) and diffraction (d-f) of speckle intensity distributions. (a) Linear crossection of the detected speckle pattern, as a function of diffusion time. (b-c) Corresponding autocorrelation function and its 1/e width squared (red circles), as well as the width squared of the autocorrelation for non-diffusive speckles (purple squares). Dashed lines present linear fit to the data. (d) Linear crossection of the detected speckle pattern, as a function of propagation distance. (e-f) Corresponding autocorrelation function and its 1/e width squared (red solid line), together with a linear fit in the linear regime $z\gg z_\text{VCZ}$ (dashed blue line). Dashed vertical line denote the propagation distance $z=z_\text{VCZ}$. The crossections and widths were obtained by considering the radial mean of the autocorrelation.}
	\label{fig:Experimental_data}
\end{figure}

Figure~\ref{fig:Experimental_data}(a) shows a 1D crossection of a 2D speckle pattern as a function of diffusion time, while normalizing the total intensity distribution at every moment, so as to account for the dissipation of the field under diffusion. As evident, the speckles grow in size with diffusion time $\tau$. This is more clearly seen in Fig.~\ref{fig:Experimental_data}(b) which presents the autocorrelation of the speckle pattern as a function of $\tau$, and in Fig.~\ref{fig:Experimental_data}(c) (red circles) which shows the width $w$ of the autocorrelation versus $\tau$. We observe that the area of the autocorrelation function grows linearly with time, i.e. $w\propto\tau^{1/2}$.

It is well known that speckles can be propagation-invariant under diffraction, if they are made by a random superposition of of Bessel beams~\cite{durnin1987diffraction, uno1995speckle, turunen1991propagation}. As we will show in detail in a future publication, Bessel beams are also invariant under diffusion, and therefore a random superposition of Bessel beams with the same radial frequency would result in a diffusion-invariant speckle field~\cite{DiffusionFree}. Consequently, the spatial coherence of such a speckle field would be diffusion-invariant as well, as demonstrated in Fig.~\ref{fig:Experimental_data}(c) (purple squares); the size of these speckles is preserved and does not increase significantly with diffusion time~\cite{Supplementary}. 

Although the specific case of such random superpositions of Bessel beams result in speckle patterns that are both diffusion and diffraction invariant, generally, there are great differences between the diffusion and diffraction of speckles. To show this explicitly, we also measured the free-space optical propagation of Gaussian speckles ~\cite{Supplementary}. The results of these measurements are shown in Figs.~\ref{fig:Experimental_data}(d-f). As evident, there are two regimes of propagation distances: Near the source at the deep Fresnel region, the size of a typical speckle is constant;
and far from the source at the VCZ region, the speckles start pulling apart from one another, and the size of a typical speckle grain gradually grows. In diffusion, on the other hand, the speckles continuously expand with diffusion time, and the relative intensity of small speckle grains decreases while the large speckle grains ``take over". Consequently, the area of the coherence region continuously grows with diffusion time.
 
\begin{figure}[htb]
    \centering
    \includegraphics{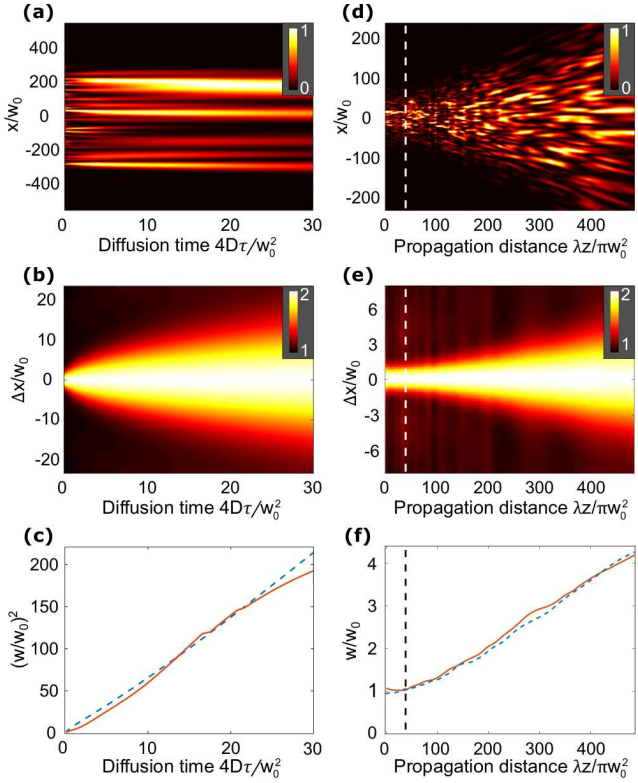}
	\caption{Numeric calculations comparing between diffusion (a-c) and diffraction (d-f) of speckle intensity distributions, and of their spatial coherence. (a) Calculated intensity distribution of a speckle pattern as a function of diffusion time. (b-c) Corresponding autocorrelation function and its 1/e width squared (solid red line). (d) Calculated intensity distribution of a speckle pattern as a function of propagation distance. (e-f) Corresponding autocorrelation function and its 1/e width squared (solid red line). The calculated intensity correlations of many such speckle realization (ensemble average) is presented in (c) and (f) as well (dashed blue line). The crossections and widths were obtained by considering the radial mean of the autocorrelation.}
	\label{fig:simulation}
\end{figure}

To validate our results, we ran numerical simulations comparing between diffraction and diffusion of speckle fields. The calculations were performed by propagating an initial speckle field using the propagators shown in Table~\ref{tab:propagators}. Figure~\ref{fig:simulation} compares between diffusion and diffraction, and qualitatively agrees with the experimental results of Fig.~\ref{fig:Experimental_data}. Furthermore, Figs.~\ref{fig:simulation}(c,f) compare the width of the autocorrelation of a single speckle pattern, to the equivalent width of coherence calculated by averaging many uncorrelated speckle patterns. As evident, the two methods are equivalent, as expected.  

\begin{table}
    \centering
    \begin{tabular}{l l l}
    \hline
     &\multicolumn{1}{c}{Diffraction} &\multicolumn{1}{c}{Diffusion}  \\ \hline
    Field & $H_E^\text{dr}=e^{-i\lambda z q^2/4\pi}$ & $H_E^\text{du}=e^{-D\tau q^2}$ \\ 
    Coherence & $H_G^\text{dr}=e^{-i\lambda z\bar{\mathbf{q}}\cdot\Delta\mathbf{q}/2\pi}$ & $H_G^\text{du}=e^{-D\tau\left(\bar{q}^2+\Delta q^2\right)}$ \\ \hline
  \end{tabular}
    \caption{Diffusion propagators derived in the paper, compared to well-known diffraction propagators.}
    \label{tab:propagators}
\end{table}
{\it Discussion and theoretical analysis.---}
We now establish the theoretical framework needed to explain the experimental and numerical results. Due to the strong relation between speckle theory and the theory of partial spatial coherence~\cite{goodman1979role, pedersen1982intensity, goodman2007speckle}, we explain the results of diffusion of speckles as diffusion of spatial coherence. We therefore begin with the familiar formalism of \textit{diffraction} of partially coherent beams, and then extend this formalism to \textit{diffusion} of such beams.  

Consider an extended quasi-monochromatic pseudothermal source, which generates a complex field amplitude $E(\mathbf{r}, z)$ at axial distance $z$ from the source and at a 2D transverse coordinate $\mathbf{r}$. The field spatial correlations are given by the mutual intensity, 
\begin{equation}
\label{eq:MutualIntensity}
    G^{(1)}\left(\mathbf{r}_1, \mathbf{r}_2; z\right)=\langle E^*\left(\mathbf{r}_1;z\right)E\left(\mathbf{r}_2; z\right)\rangle,
\end{equation}
where $E^*$ is the complex conjugated of $E$, and $\langle\cdot\rangle$ denotes ensemble average. The first order intensity correlation $G^{(2)}(\mathbf{r}_1, \mathbf{r}_2; z) = \langle I\left(\mathbf{r}_1; z\right)I\left( \mathbf{r}_2; z\right)\rangle$ is related to $G^{(1)}(\mathbf{r}_1, \mathbf{r}_2; z)$ by the Siegert relation, 
\begin{equation}
    G^{(2)}(\mathbf{r}_1, \mathbf{r}_2; z) = \langle I\left( \mathbf{r}_1; z\right)\rangle \langle I\left( \mathbf{r}_2; z\right)\rangle + |G^{(1)}(\mathbf{r}_1, \mathbf{r}_2; z)|^2.
    \label{eq:Siegert}
\end{equation}
For a random speckle field which contains a sufficiently large number of speckles, $G^{(2)}$ can  be equivalently measured by taking the autocorrelation of a single speckle field~\cite{goodman1979role, pedersen1982intensity}.

In the following we consider a planar quasi-homogeneous source, namely a source whose area $L^2$ is very large compared to the coherence area $l_\text{c}^2$ on the source, and any variations in intensity on the source occur on size scales that are of order $L^2$. If the two scales are well separated, $L\gg l_\text{c}$, the mutual intensity $G_0^{(1)}$ at the plain of the source $z=0$ can then be factorized, 
\begin{equation}
    G^{(1)}_0(\mathbf{r}_1, \mathbf{r}_2)=I_0(\Bar{\mathbf{r}})\mu_0(\Delta\mathbf{r}),     
\end{equation}
where $\bar{\mathbf{r}}\equiv (\mathbf{r}_1+\mathbf{r}_2)/\sqrt{2}$ and $\Delta\mathbf{r}\equiv(\mathbf{r}_2-\mathbf{r}_1)/\sqrt{2}$. The term $I_0(\bar{\mathbf{r}})$ signifies variations on the scale of $L$, while $\mu_0(\Delta\mathbf{r})$ signifies variations on the scale of $l_\text{c}$. Under this assumption, it is convenient to describe the mutual intensity at some plane $z$ away from the source in Fourier space, where we use the propagator $H_E^\text{du}$ and obtain ~\cite{Supplementary} 
\begin{equation}
    \begin{split}
        \Tilde{G}^{(1)}\left(\bar{\mathbf{q}}, \Delta\mathbf{q}; z\right)=\Tilde{G}_0^{(1)}\left(\bar{\mathbf{q}}, \Delta\mathbf{q}\right) e^{i\lambda z\bar{\mathbf{q}}\cdot \Delta\mathbf{q}/2\pi},
        \label{eq:G_Fresnel}
    \end{split}
\end{equation}
where $\Tilde{G}^{(1)}$ and $\Tilde{G}_0^{(1)}$ are the Fourier transforms of $G^{(1)}$ and $G_0^{(1)}$, respectively, and $\bar{\mathbf{q}}\equiv (\mathbf{q}_1+\mathbf{q}_2)/\sqrt{2}$ and $\Delta\mathbf{q}\equiv(\mathbf{q}_2-\mathbf{q}_1)/\sqrt{2}$ are the Fourier coefficients. The evolution of the mutual intensity $G^{(1)}$ with propagation distance $z$ can therefore be calculated using the propagator $H_G^{dr}=e^{i\lambda z\bar{\mathbf{q}}\cdot \Delta\mathbf{q}/2\pi}$. In the near vicinity of the source $z\ll Ll_\text{c}/\lambda\equiv z_\text{VCZ}$, the propagator $H_G^{dr}$ is negligible, resulting in $G^{(1)}\left(\bar{\mathbf{r}}, \Delta\mathbf{r}; z\right)=G^{(1)}_0\left(\bar{\mathbf{r}}, \Delta\mathbf{r}\right)$. Therefore, in this region, often referred to as the deep Fresnel region~\cite{cerbino2007correlations}, the mutual intensity is constant and does not vary with propagation distance $z$~\cite{ohtsuka1981non, turunen1991propagation, giglio2000space}. Indeed, the experimental data of Figs.~\ref{fig:Experimental_data}(d-f) show that in this region, the spatial coherence of the speckles is constant, and does not vary with propagation distance. 

Following Gatti~\cite{gatti2008three}, far from the source $z\gg z_\text{VCZ}$, Eq.~(\ref{eq:G_Fresnel}) can be solved to yield the generalized VCZ theorem,~\cite{goodman2015statistical, goodman2007speckle}
\begin{equation}
    \begin{split}
        G^{(1)}\left(\bar{\mathbf{r}}, \Delta\mathbf{r}; z\right) &= 
        \left(\frac{2\pi}{\lambda z}\right)^2  \Tilde{G}_0^{(1)}\left(\frac{2\pi}{\lambda z}\bar{\mathbf{r}}, \frac{2\pi}{\lambda z}\Delta\mathbf{r}\right)e^{\left(2\pi i/\lambda z\right)\bar{r}\Delta r}.
        \label{eq:VCZ}
    \end{split}
\end{equation}
which expresses a Fourier relation between the mutual intensity $G^{(1)}$ at the plane of the source $z=0$ and far from it, in the Fresnel region $z\gg z_\text{VCZ}$. 

Recall that $\Tilde{G}_0^{(1)}(\bar{\mathbf{q}}, \Delta\mathbf{q})=\Tilde{I}_0(\Delta\mathbf{q})\cdot\Tilde{\mu}_0(\bar{\mathbf{q}})$, where $\Tilde{\mu}_0$ and $\Tilde{I}_0$ are the Fourier transforms of $\mu_0$ and $I_0$, respectively. For $L\gg l_\text{c}$ and $z\gg z_\text{VCZ}$, the function $\Tilde{\mu}_0$ varies very slowly as compared to $\Tilde{I}_0$, and therefore $\Tilde{\mu}_0$ is approximately constant. In this case the absolute value of the mutual intensity $|G^{(1)}|$ depends only on $\Tilde{I}_0$, and therefore the coherence region grows as $\lambda z/L$. Indeed, Fig.~\ref{fig:Experimental_data}(f) shows a divergence slope of $1.30\cdot 10^{-3}\pm0.04\cdot10^{-3}$, which agrees with the expected slope of $1.33\cdot 10^{-3}$.

We now turn to derive equivalent expressions to describe diffusion of partially coherent beams. As before, we begin with Eq.~(\ref{fig:Experimental_data}), but now we express the fields using the diffusion field propagator $H_E^\text{du}$, yielding in Fourier space~\cite{Supplementary}
\begin{equation}
    \begin{split}
        \Tilde{G}^{(1)}\left(\bar{\mathbf{q}}, \Delta\mathbf{q}; \tau\right) &= 
        \Tilde{G}_0^{(1)}\left(\bar{\mathbf{q}}, \Delta\mathbf{q}\right)e^{-D\tau\left(\bar{q}^2+\Delta q^2\right)}\\
        &=\left[\Tilde{\mu}_0(\bar{\mathbf{q}})e^{-D\tau\bar{q}^2}\right]\cdot\left[\Tilde{I}_0(\Delta\mathbf{q}) e^{-D\tau\Delta q^2}\right].
        \label{eq:GDiffusion}
    \end{split}
\end{equation}	 
This equation is the diffusion analogue of Eq.~(\ref{eq:G_Fresnel}), as they both describe the evolution of the spatial coherence in Fourier space. In diffraction, the local variations described by $\Tilde{\mu}_0$ and the global variations described by $\Tilde{I}_0$ are ``mixed'' by the propagator $H_G^\text{dr}=e^{i\lambda z\bar{\mathbf{q}}\cdot\Delta\mathbf{q}/2\pi}$. As $\Tilde{\mu}_0$ represents local variations and $\Tilde{I}_0$ represents global variations, the propagator $H_G^\text{dr}$ couples between the local and global coherence properties of the source with propagation distance $z$. In contrast, in diffusion the propagator of the mutual intensity $H_G^\text{du}=e^{-D\tau\left(\bar{\mathbf{q}}\cdot\Delta\mathbf{r}+\Delta\mathbf{q}\cdot\bar{\mathbf{r}}\right)}$ can be factorized, resulting in the two independent terms expressing the mutual intensity in Eq.~(\ref{eq:GDiffusion}) for any diffusion time $\tau$. Therefore, the local and global coherence features undergo diffusion without ``mixing''. 

For $L\gg l_\text{c}$, Eq.~(\ref{eq:GDiffusion}) implies that the width along the relative coordinate $\Delta\mathbf{r}$ of the mutual intensity $G^{(1)}\left(\bar{\mathbf{r}}, \Delta\mathbf{r}; \tau\right)$ at any point in time depends mainly on the width of the coherence region on the surface of the source $l_\text{c}$, and only weakly on the width of the source itself $L$, even far from the source. This is very different from the behavior of the mutual intensity under diffraction, where far from the source, the spatial coherence depends on  the shape and size of the source. This leads to a significant difference between diffusion and diffraction in a Michelson or Hanbury Brown and Twiss (HBT) type of interferometers. In diffraction, a HBT interferometer can be used to measure the size of a distant spatially incoherent object, but cannot be used to retrieve information regarding the original size of coherent regions on the source. However, in diffusion the picture is reversed (but is similar to the case of diffraction in the deep Fresnel region): Measuring the spatial coherence indicates the size of coherence regions at a distant source (albeit with accuracy that decays with diffusion time), and will supply little information regarding size of the source itself. 

For a Gaussian speckle field with Gaussian envelope, $\mu(\Delta\mathbf{r})=e^{-\Delta\mathbf{r}^2/l_\text{c}^2}$, $I(\bar{\mathbf{r}})=I_0e^{-\bar{\mathbf{r}}}$, we obtain  
\begin{equation}
    G^{(1)}\left(\bar{\mathbf{r}}, \Delta\mathbf{r}; \tau\right) =
  \exp\left(-\frac{\Delta\mathbf{r}^2}{l_\text{c}^2+4D\tau}\right)
    \cdot
    \exp\left(-\frac{\bar{\mathbf{r}}^2}{L^2  +4\tau}\right).
\end{equation}
In the limit $L\gg l_\text{c}$, the expected width squared is $w^2\approx l_\text{c}^2+4D\tau$. Indeed, Fig.~\ref{fig:Experimental_data}(c) shows a divergence slope of \SI[separate-uncertainty = true, multi-part-units=single]{40.5(4)}{cm\squared\per\second}, which agrees with an independent measurement of the diffusion coefficient, $4D=\SI[separate-uncertainty = true, multi-part-units=single]{38.8(22)}{cm\squared\per\second}$~\cite{Supplementary}. 

The number of speckles $N$ at the plane of the source is  estimated by $N\sim(L/l_\text{c})^2$. In diffraction, the number of speckles is conserved, and does not vary even after long propagation distances. In diffusion, however, the number of speckles reduces with diffusion time, and the expected number of speckles after diffusion time $\tau$ is $N\sim(L^2+4D\tau)/(l_\text{c}^2+4D\tau)$. 

{\it Concluding remarks.---}
We analyzed diffusion of partially coherent complex fields, and compared between diffusion and diffraction of the spatial coherence. We showed, both in theory and experiment, that the complex field and the spatial coherence of partially coherent beams both undergo diffusion in a similar manner. While diffraction of partially coherent beams behaves differently in the the deep Fresnel region and the VCZ region, diffusion of the partial coherence behaves the same in all space. As we showed, diffusion of partial coherence leads to a diffusion-analogue of the classical Michelson or HBT interferometers. In diffusion, the source\textsc{\char13}s boundary has little effect on the spatial coherence, and  measuring the spatial coherence far from the source can be considered as a measurement of the original region of coherence at the source.  


The work presented here extended concepts and theorems from statistical optics to the field of coherent diffusion. While we focused here on polariton diffusion, our analysis is general, and provides a first step in applying the VCZ theory and HBT interferometry to various diffusive systems, such as astronomical stellar atmospheres~\cite{ryabchikova2014diffusion}, and imaging through turbulent or complex scattering media~\cite{mosk2012controlling, rotter2017light}.

\begin{acknowledgments}
The research presented here was supported by the Israel Science Foundation (ISF), the Israel-US Binational Science Foundation (BSF) and by the Pazi foundation.  
\end{acknowledgments}

\bibliographystyle{apsrev4-1}
\bibliography{main.bib}

\end{document}



\heading{Supplementary Material for\\Coherent diffusion of partial spatial coherence}
\begin{center} Ronen Chriki, Slava Smartsev, David Eger, Ofer Firstenberg \\ and Nir Davidson\end{center}
\begin{center}
\footnotesize{\textit{Department of Physics of Complex Systems, Weizmann Institute of Science, \\ Rehovot 7610001, Israel}} 
\end{center}


\subsection*{Experimental arrangement for diffusion of speckles}
We used \Rb vapor with \SI{10}{\torr} buffer gas of $\text{N}_2$, heated to \SI{65}{\celsius} and placed in a \SI{7.5}{\cm} long cell. The cell is held inside a three layered shield to isolate it from external magnetic field, and a weak \SI{50}{\milli\gauss} longitudinal magnetic field assures that the spectator ground states are far from the Raman resonance. 

An amplified \SI{795}{\nm} diode laser with (one photon) linewidth of \SI{1}{\mega\hertz} is split into three beams. As mentioned in the manuscript, two of these beams, denoted as control beams, overlap spatially in the area of the vapor cell, but are separated by a slight angle of $\theta\approx\SI{10}{\milli\radian}$. The third beam, denoted as the probe, is modulated at $\sim\SI{6.8}{\giga\hertz}$, and illuminated onto a spatial light modulator (SLM) which controls its transverse complex profile. The two-photon detuning $\Delta_{\text{2p}}$ is scanned by varying the modulation frequency of the probe by $\pm\SI{15}{\kilo\hertz}$ at most. The probe is oriented such that it propagates along one of the control beams inside the vapor cell. The purpose of the three beams is to generate a fourth signal beam along the path of the second control. The diameters of all three beams is $\gtrsim\SI{8}{\milli\meter}$. After the cell, the generated signal is spatially separated from the first control and from the probe. In order to detect the signal beam, the second probe is filtered with etalons, and the signal is imaged onto a CCD camera.

\subsection*{Experimental arrangement for diffraction of speckles}
\begin{figure}
    \centering
    \includegraphics{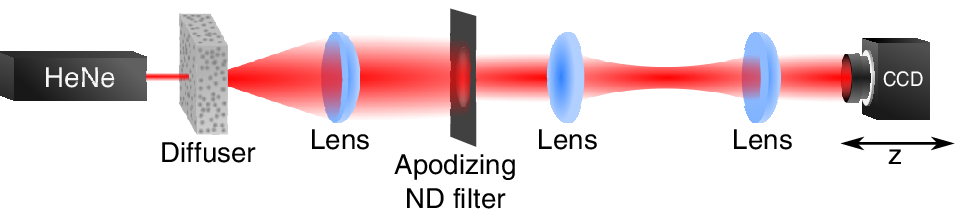}
    \caption{Experimental arrangement for diffraction of speckles.}
    \label{fig:diffraction_exparrangement}
\end{figure}

The experimental arrangement used to characterize speckle dynamics under diffraction is rather simple. As shown in Fig.~\ref{fig:diffraction_exparrangement}, it is comprised of HeNe single mode laser that illuminates a diffuser (Newport \SI{5}{\degree} light shaping diffuser), and then collimated by a lens of focal distance $f=\SI{20}{\cm}$. The resulted speckle pattern is truncated by an apodized neutral density filter, and then imaged onto a CCD camera, which can shift axially using a computerized translation stage. In our experiments, the camera was moved axially in steps of $\SI{0.5}{\mm}$.

\subsection*{Independent measurement of the diffusion constant \textit{D}}
\begin{figure}
    \centering
    \includegraphics{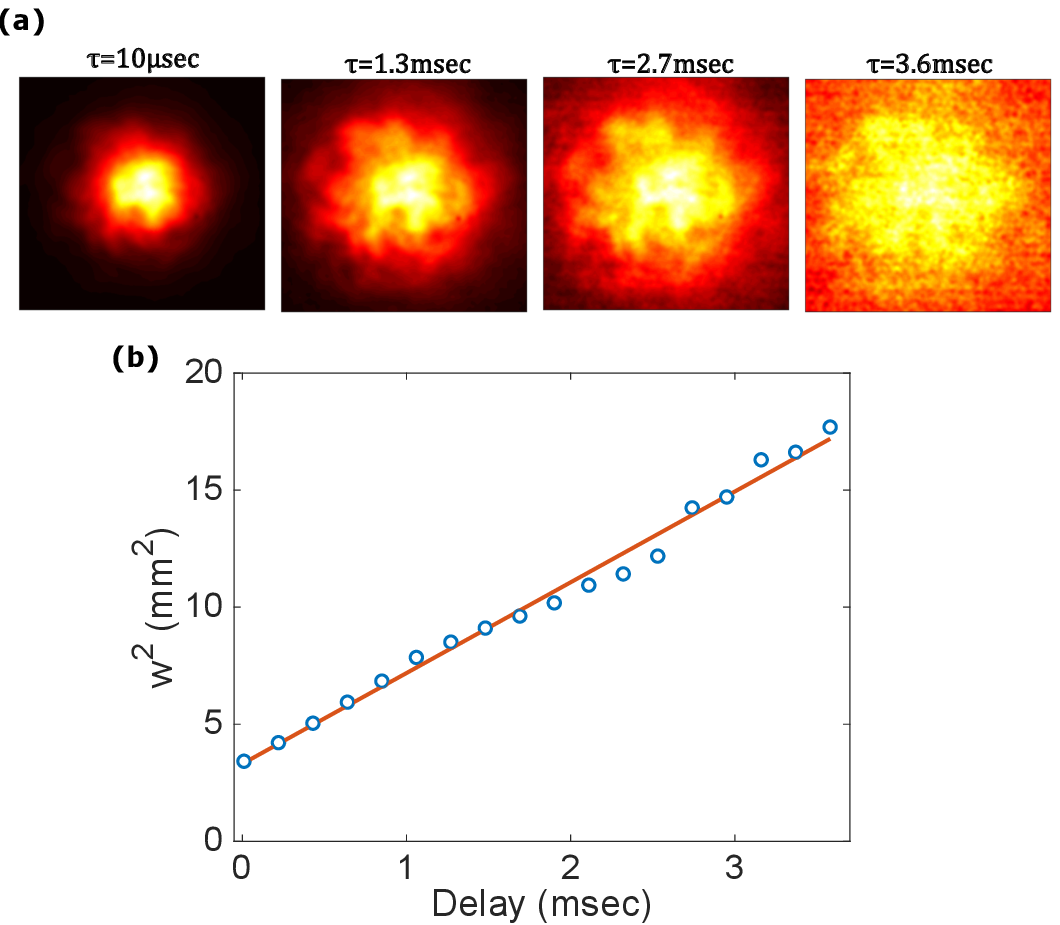}
    \caption{Independent measurement of the diffusion coefficient. (a) Detected intensity distributions after temporal delays of $\tau=\SI{10}{\micro\second}$, $\tau=\SI{1.3}{\milli\second}$, $\tau=\SI{2.7}{\milli\second}$ and $\tau=\SI{3.6}{\milli\second}$. (b) Radial mean of the detected beam's width squared, as a function of delay time.}
    \label{fig:storage}
\end{figure}

The diffusion constant was measured independently using a storage and retrieval experiment~\cite{vudyasetu2008storage, firstenberg2010self, smartsev2017continuous}, where a broad pump and a narrow Gaussian profiled probe are used to store a narrow Gaussian onto the \Rb vapor. The beams are then turned off for duration $\tau$, after which the second pump is turned on, and the signal is retrieved.  
Figure~\ref{fig:storage}(a) presents intensity distribution of the detected signal, for various delay times. As evident, the size of the input beam grows with delay. Figure~\ref{fig:storage}(b) shows the radial mean of the beam width squared, as a function of delay. From the slope of this curve we can extract the diffusion constant, which was found to be $D=\SI[separate-uncertainty = true, multi-part-units=single]{9.7(5)}{cm\squared\per\second}$. This result agrees with the calculated values~\cite{ma2009modification, ishikawa2000diffusion}.

\subsection*{Diffusion and induced diffraction of spatial coherence}
For convenience, we denote the $\ket{\pm}$ states in the rotating frame, $\ket{\pm}=\left(\ket{3}\pm\ket{2}\right)/\sqrt{2}$, and the corresponding normal modes $E_+$ and $E_-$. Accordingly, we express the normal modes immediately at the exit of the \Rb medium as $E_\pm^{out}=g_\pm E_\pm^{in}$, where $g_-=e^{-S}$ describes regular (one-photon) absorption and $g_+ = e^{-S\left(1-f\right)}$ describes electromagnetically induced transparency (EIT); $S$ and $f$ being the Lorentzians associated with the one and two photon resonances,
\begin{equation}
\label{eq:define_SandF}
        S = d\frac{\gamma_{\text{1p}}}{\gamma_{\text{1p}}-i\Delta_{\text{1p}}},\qquad
        f = \eta_{act}\frac{\Gamma}{\gamma_{\text{2p}}+\Gamma-i\Delta_{\text{2p}}}.
\end{equation}
Here, $d$ is half the resonant optical depth for the probe, $\gamma_{\text{1p}}$ and $\gamma_{\text{2p}}$ are the one and two photon decoherence rates, $\Delta_{\text{1p}}$ and $\Delta_{\text{2p}}$ are the one and two photon frequency detunings, and $\Gamma$ denotes the power broadening $\Gamma=\Omega^2/\left(\gamma_{\text{1p}}-i\Delta_{\text{1p}}\right)$, $\Omega$ being the Rabi frequency of the control beams. The prefactor $0\leq\eta_{act}\leq1$ is the fraction of atoms in the lower hyperfine manifold that populated the $m=0$ (lower $\ket{1}$) state.  

For a uniform incoming probe field $E_\text{in}$, the generated signal in the limit of weak EIT $|Sf|\ll 1$ is given by
\begin{equation}
     \frac{E_s}{E_\text{in}} = \frac{1}{2}\left(g_+ - g_-\right)\approx{\frac{1}{2}Sfe^{-S}},
    \label{eq:generated1}
\end{equation}
where $g_+$ and $g_-$ are the transmission amplitudes with and without EIT, respectively, and $S$ and $f$ are the complex Lorentzians associated with the one and two-photon detunings (Eq. 2). The generated signal has a group delay of
\begin{equation}
        \tau=\frac{\partial}{\partial\Delta_{\text{2p}}}\left[\log\left(\frac{g_+-g_-}{2}\right)\right] =  \frac{\gamma_{\text{2p}}+\Gamma-i\Delta_{\text{2p}}}{\left(\gamma_{\text{2p}}+\Gamma\right)^2+\left(\Delta_{\text{2p}}\right)^2}.
    \label{eq:diffusion_time1}
\end{equation}
Eq.~\ref{eq:diffusion_time1} indicates that $\tau$ has both real and imaginary components, $\tau = \tau_\text{du}+i\tau_\text{dr}$. The real term $\tau_\text{du}$ corresponds to real diffusion, and the imaginary term $\tau_\text{dr}$ corresponds to motional-induced diffraction. In the experiment, we vary the diffusion time $\tau$ by changing the two photon detuning $\Delta_{\text{2p}}$. Notice that for a given $\gamma_{\text{2p}}+\Gamma = const$, the maximal diffusion delay time $\text{Re}\left[\tau\right]$ is obtained for $\Delta_{\text{2p}}=0$. Accordingly, we denote the maximal diffusion time $\tau_\infty\equiv1/(\gamma_{\text{2p}}+\Gamma)$, resulting in 
\begin{equation}
        \frac{\tau}{\tau_\infty}= \frac{1-i\Delta_{\text{2p}}\tau_\infty}{1+\left(\Delta_{\text{2p}}\tau_\infty\right)^2}.
    \label{eq:diffusion_time2}
\end{equation}

To estimate the effect of the induced diffraction, we consider the evolution of a Gaussian beam of initial waist $w_0$. In analogy to the Rayleigh range in optics, we define a Rayleigh duration $\tau_R$ after which the the beam expands by a factor of $\sqrt{2}$ due to diffusion. The width of the beam after some diffusion time $\tau$ is simply 
\begin{equation}
    w(\tau)^2=w_0^2+4D\tau,
\end{equation}
and therefore $\tau_R=\frac{w_0^2}{4D}$. The induced diffraction becomes significant if the duration $\tau_\text{dr}$ is large as compared to the effect of diffusion,
\begin{equation}
    |\tau_\text{dr}|>\tau_\text{R}+\tau_\text{du}.
\end{equation}

Plugging Eq.~(\ref{eq:diffusion_time2}) and solving for $\Delta_{\text{2p}}\tau_\infty$, we find the minimal two-photon detuning needed to obtain substantial induced diffraction. Using the parameters used in our experiment, considerable induced diffraction is expected for $140>|\tau_\infty\Delta_{\text{2p}}|>3.3$. Therefore, the effect of induced diffraction can be safely neglected in the regime studied in the manuscript ($3>\tau_\infty\Delta_{\text{2p}}>0$). We have verified the above analysis using a numerical simulation, comparing diffusion of a speckle field with and without induced diffraction, $\tau=\tau_\text{du}+i\tau_\text{dr}$ and $\tau=\tau_\text{du}$, respectively.

\begin{figure}
    \centering
    \includegraphics{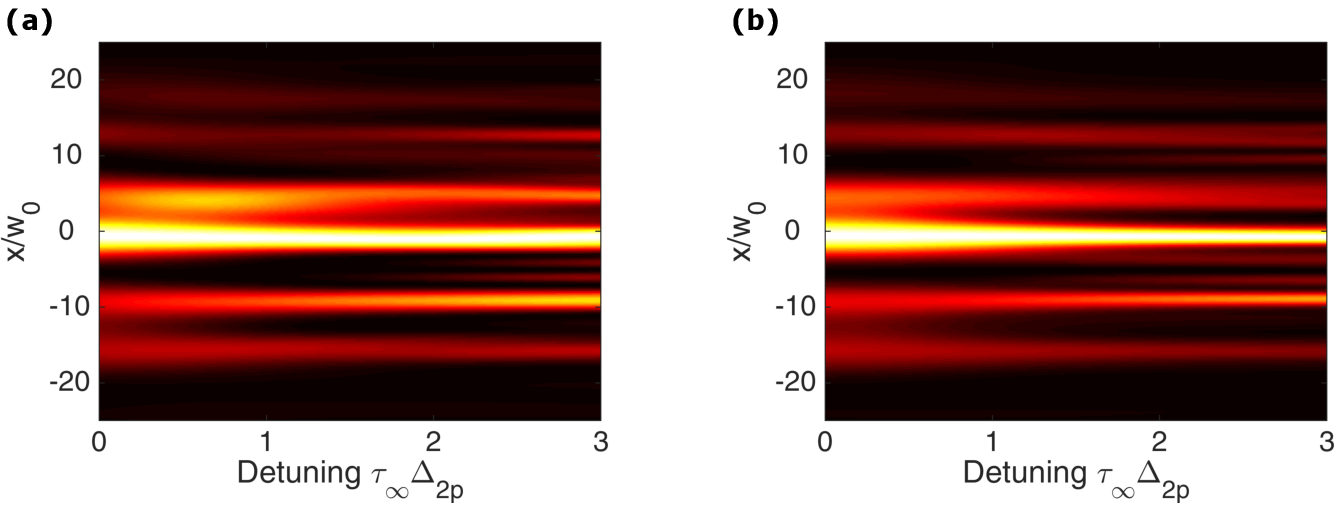}
    \caption{Calculated results for diffusion and induced diffraction. (a) Calculated intensity distribution as a function of two photon detuning (equivalent to diffusion time), without  the term of induced diffraction. (b) Calculated intensity distribution as a function of two photon detuning (equivalent to diffusion time), considering also the term of induced diffraction. The parameters taken for the simulation are similar to those in experiment.}
    \label{fig:simulations_diffraction}
\end{figure}

\subsection*{Evolution of the spatial coherence - derivation}

\subsubsection*{I. Diffraction}
We begin with diffraction of partial spatial coherence, where, as in the manuscript, the mutual intensity is define as~\cite{goodman2015statistical, mandel1995optical}
\begin{equation}
    G^{(1)}\left(\mathbf{r}_1, \mathbf{r}_2; z\right) = \langle E^*\left(\mathbf{r}_1; z\right)E\left(\mathbf{r}_2; z\right)\rangle.
    \label{eq:Gdefinition}
\end{equation}
It is convenient to shift to Fourier space to describe the evolution of the mutual intensity $G^{(1)}$, where, using the Fresnel Fourier space propagator $H_E^\text{dr}=e^{-i\lambda z q^2/4\pi}$~\cite{goodman2008introduction}, we obtain
\begin{equation}
    \Tilde{G}^{(1)}\left(\mathbf{q}_1, \mathbf{q}_2; z\right) = \Tilde{G}_0^{(1)}\left(\mathbf{q}_1, \mathbf{q}_2\right) e^{-i\lambda z\left(q_2^2-q_1^2\right)/4\pi},
    \label{eq:GFourier_diffraction}
\end{equation}
where, $\Tilde{G}_0^{(1)}\left(\mathbf{q}_1, \mathbf{q}_2\right)=\langle \Tilde{E}^*\left(\mathbf{q}_1; z=0\right)\Tilde{E}\left(\mathbf{q}_2; z=0\right)\rangle$ is the Fourier transform of the mutual intensity $G^{(1)}$ at the plane of the source $z=0$.

Changing variables to $\Delta\mathbf{q}$ and $\bar{\mathbf{q}}$, Eq.~\ref{eq:GFourier_diffraction} yields
\begin{equation}
    \Tilde{G}^{(1)}\left(\bar{\mathbf{q}}, \Delta\mathbf{q}; z\right) = \Tilde{G}_0^{(1)}\left(\bar{\mathbf{q}}, \Delta\mathbf{q}\right) e^{-i\lambda z\left(\bar{\mathbf{q}}\cdot\Delta\mathbf{q}\right)/2\pi}.
    \label{G_deltaRep_diffraction}
\end{equation}
We therefore conclude the diffraction of the mutual intensity can be described by the Fourier space propagator $H_G^\text{dr}=e^{-i\lambda z\left(\bar{\mathbf{q}}\cdot\Delta\mathbf{q}\right)/2\pi}$, which mixes between the coordinate axes $\bar{\mathbf{q}}$ and $\Delta\mathbf{q}$.

\subsubsection*{II. Diffusion}
Similarly, using the diffusion propagator $H_G^\text{du}=e^{-D\tau q^2}$, the mutual intensity in Fourier space yields
\begin{equation}
    \Tilde{G}^{(1)}\left(\mathbf{q}_1, \mathbf{q}_2; \tau\right) = \Tilde{G}_0^{(1)}\left(\mathbf{q}_1, \mathbf{q}_2\right) e^{-D\tau\left(q_1^2+q_2^2\right)}.
    \label{eq:GFourier_diffusion1}
\end{equation}
Equivalently, changing variables to $\bar{\mathbf{q}}$ and $\Delta\mathbf{q}$, Eq.~\ref{eq:GFourier_diffusion1} can be expressed as
\begin{equation}
    \Tilde{G}^{(1)}\left(\bar{\mathbf{q}}, \Delta\mathbf{q}; \tau\right) = \Tilde{G}_0^{(1)}\left(\bar{\mathbf{q}}, \Delta\mathbf{q}\right) e^{-D\tau\left(\bar{\mathbf{q}}^2+\Delta\mathbf{q}^2\right)}.
    \label{eq:GFourier_diffusion2}
\end{equation}
Thus, the diffusion propagator of the mutual intensity is $H_G^\text{du}=e^{-D\tau\left(\bar{\mathbf{q}}^2+\Delta\mathbf{q}^2\right)}$. The different diffusion and diffraction propagators are summarized in table I in the manuscript.

\subsection*{Diffusion invariant speckles}
Over the past three decades, ever since the pioneering work of Durnin~\cite{durnin1987exact, durnin1987diffraction}, Bessel beams have attracted much attention, due to their unique propagation properties, where they can maintain their transverse profile for long propagation distances~\cite{mcgloin2005bessel}. Recently, Bessel beams were also shown to be immune to diffusion[ref!]. Specifically, except for global dissipation in energy, a Bessel beam that undergoes diffusion maintains its transverse profile for long temporal duration. 

This diffusion invarience of Bessel beams can be intuitively understood by considering their Fourier component representation. Bessel beams can be decomposed to a superposition of plane waves of equal energy uniformly distributed on a cone. Since the diffusion propagator $H_E^\text{du}$ depends only on the squared norm of the wavevector, all plane waves undergo the exact same dissipation. Since Bessel beams are diffusion invariant, any random superposition of Bessel beams is diffusion invariant as well. It is therefore possible to generate speckle field which would undergo minimal diffusion, as demonstrated in Fig. 2(c) of the manuscript.

\begin{figure}
    \centering
    \includegraphics{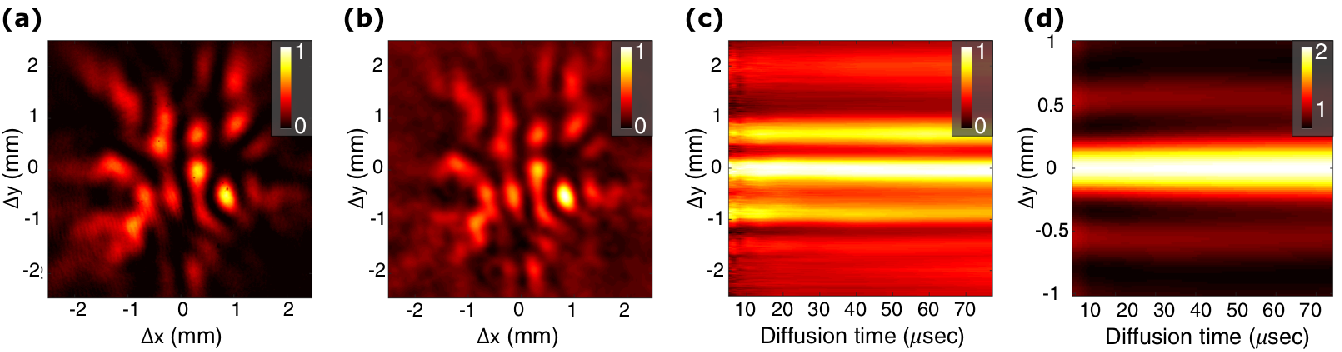}
    \caption{Diffusion invariant speckles. (a) Detected intensity distribution of the detected speckle pattern after short diffusion time ($\tau=5 \mu sec$). (b) Detected intensity distribution of the detected speckle pattern after long diffusion time ($\tau=76 \mu sec$).(c) Crossection of the intensity distribution of the detected speckle pattern, as a function of diffusion time. (d) Crossection of the calculated auto-correlation of the detected speckle pattern. 
    }
    \label{fig:diffusion_invariant_supp}
\end{figure}

In the following, we present additional experimental results for diffusion invariant speckle, which complement the data presented in the manuscript~\cite{turunen1991propagation, uno1995speckle}. Figures~\ref{fig:diffusion_invariant_supp}(a-b) show the detected intensity distribution of the generated signal after short and long diffusion time. As evident, the intensity distribution does not vary with diffusion time. This is more clearly evident in Fig.~\ref{fig:diffusion_invariant_supp}(c) which presents a vertical crossection of the speckle pattern, as a function of diffusion time. Since the field itself is invariant to diffusion, it is only expected that the statistical properties of the speckle field should be invariant as well. Indeed, Fig.~\ref{fig:diffusion_invariant_supp}(d) shows that the calculated speckle auto-correlation maintains its profile and does not vary with diffusion time.   

\bibliographystyle{apsrev4-1}
\bibliography{main.bib}